\documentclass{emulateapj}
\usepackage{multirow}
\usepackage{subfigure}
\def\msun{$M_{\odot}$} 
 
\def\rsun{$R_{\odot}$}

\tabletypesize{\tiny}
\shortauthors{Lin et al.}
\begin{document}

\title{Multiwavelength Follow-up of the Hyperluminous
  Intermediate-mass Black Hole Candidate 3XMM~J215022.4$-$055108}

\author{Dacheng Lin\altaffilmark{1},  Jay Strader\altaffilmark{2},  Aaron
  J. Romanowsky\altaffilmark{3,4},  Jimmy A. Irwin\altaffilmark{5}, Olivier Godet\altaffilmark{6,7}, Didier Barret\altaffilmark{6,7}, Natalie A. Webb\altaffilmark{6,7}, Jeroen Homan\altaffilmark{8,9},
Ronald A. Remillard\altaffilmark{10}}
\altaffiltext{1}{Space Science Center, University of New Hampshire,
  Durham, NH 03824, USA, email: dacheng.lin@unh.edu}
\altaffiltext{2}{Center for Data Intensive and Time Domain Astronomy,
  Department of Physics and Astronomy, Michigan State University, 567
  Wilson Road, East Lansing, MI 48824, USA}
 \altaffiltext{3}{Department of Physics and Astronomy, San Jos\'{e} State University, One Washington Square, San Jos\'{e}, CA 95192, USA}
\altaffiltext{4}{University of California Observatories, 1156 High
  Street, Santa Cruz, CA 95064, USA}
\altaffiltext{5}{Department of Physics and Astronomy, University of Alabama, Box 870324, Tuscaloosa, AL 35487, USA}
\altaffiltext{6}{CNRS, IRAP, 9 avenue du Colonel Roche, BP 44346, F-31028 Toulouse Cedex 4, France}
\altaffiltext{7}{Universit\'{e} de Toulouse, UPS-OMP, IRAP, Toulouse, France}
\altaffiltext{8}{Eureka Scientific, Inc., 2452 Delmer Street, Oakland, California 94602, USA}
\altaffiltext{9}{SRON, Netherlands Institute for Space Research,
  Sorbonnelaan 2, 3584 CA Utrecht, The Netherlands}
\altaffiltext{10}{MIT Kavli Institute for Astrophysics and Space Research, MIT, 70 Vassar Street, Cambridge, MA 02139-4307, USA}

\begin{abstract}
We recently discovered the X-ray/optical outbursting source 
3XMM~J215022.4$-$055108. It was best explained as the tidal disruption of
a star by an intermediate-mass
black hole of mass of a few tens of thousand solar masses in a massive
star cluster at the outskirts of a large barred
lenticular galaxy at $D_L=247$ Mpc. However, we could not completely
rule out a Galactic cooling neutron star as an alternative explanation
for the source. In order to further pin down the nature of the source, we
have obtained new multiwavelength observations by \emph{XMM-Newton}
and \emph{Hubble Space Telescope} (\emph{HST}). The optical
counterpart to the source in the new \emph{HST}
image is marginally resolved, which rules out the Galactic cooling
neutron star explanation for the source and suggests a star cluster of
half-light radius $\sim$27 pc.  The new \emph{XMM-Newton} observation indicates that the
luminosity was decaying as expected for a tidal disruption event and that the
disk was still in the thermal state with a super-soft X-ray
spectrum. Therefore, the new observations confirm
the source as one of the best intermediate-mass
black hole candidates. 
\end{abstract}


\section{INTRODUCTION}
\label{sec:intro}
There has been strong evidence for the existence of stellar-mass black
holes (BHs, mass $\sim10$ \msun) from dynamical measurements
\citep{remc2006} and gravitational wave detections
\citep{ababab2016}. The evidence for the existence of supermassive
BHs (SMBHs, mass $\sim10^6$--$10^{10}$ \msun) at the centers of
massive galaxies is also very compelling \citep{kori1995,gieitr2009,
  grabam2018,evakal2019}. However, intermediate-mass BHs of mass
$\sim$100--$10^5$ \msun\ are still observationally elusive despite
the long-term search \citep[see][for a recent review]{grstho2019}. Some candidates were found from a variety of systems, including dwarf
galaxies \citep[e.g.,][]{dowayu2007,barega2015,chkazo2018}, globular
clusters \citep[e.g.,][]{irbrbr2010,kibalo2017,nogebe2008,pestly2017},
and hyperluminous off-nuclear X-ray sources \citep[HLXs, X-ray luminosity
$L_\mathrm{X}\ge10^{41}$
erg~s$^{-1}$,][]{faweba2009,wecsle2012,licawe2016}. Confirming the
IMBH nature of these candidates is non-trivial, as there is no well
accepted feasible method to weigh these BHs. The inference of the
IMBHs in globular clusters depends on the models/methods used to infer
their presence and is typically called
into question in follow-up studies \citep[e.g.,][]{marihe2019,bahesw2019}.
HLXs are interesting IMBH candidates, but it is important to rule
out background AGNs \citep{surogl2015} or even accreting neutron
stars \citep{isbest2017}. 

We reported our discovery of a new HLX candidate
3XMM~J215022.4$-$055108 (J2150$-$0551 hereafter) in \citet[][Lin18
hereafter]{listca2018}.  The source exhibited a prolonged X-ray
outburst of peak X-ray flux of $\sim$$10^{-12}$ erg s$^{-1}$
cm$^{-2}$, lasting for more than a decade and X-ray spectra soft and purely thermal. The outburst was also detected in
the optical. The source is located at the outskirts of a large barred lenticular
galaxy (Gal1 hereafter) at $z=0.055$, and we identified a faint optical counterpart based on the
positional coincidence and the correlated optical/X-ray variability. The
most promising explanation for the source is that it is an IMBH in an
off-center star cluster with the X-ray/optical outburst (peak X-ray
luminosity $\sim7\times10^{42}$ erg~s$^{-1}$) due to a tidal
disruption event (TDE), in which a star having
a close encounter with the BH was tidally disrupted and
subsequently accreted, producing the multiwavelength flare
\citep[see][for a recent review]{ko2015}. We measured the BH mass to be $\sim5\times10^4$ \msun,
based on the fit to the X-ray spectra, which we assumed to be in the
thermal state during the decay. The thermal state identification was
supported by the fact that the X-ray spectra can be described well with a
standard thin disk, whose temperature and luminosity approximately
followed the $L\propto T^4$ relation. The event was later modeled in
detail by \citet{chsh2018}, who inferred a main-sequence disrupted star of mass 0.33 \msun\ and radius 0.41 \rsun. 

The star cluster was not clearly resolved in a \emph{Hubble Space
Telescope} (\emph{HST}) Advanced Camera for Surveys (ACS) Wide Field
Camera (WFC) F775W image in 2003 before the outburst. This is probably
due to the low signal-to-noise ratio of the source in this image, in
which the source fell into the CCD gap in two of four exposures. Based
on the fit to the broad-band quiescent photometry, we inferred the
stellar mass of the cluster to be
$\sim$$10^7$ \msun, making it either a massive globular cluster or an 
ultra-compact dwarf (UCD) galaxy, which has physical properties
intermediate between classical globular clusters and galaxies and is
often explained as a remnant nucleus of a tidally stripped dwarf galaxy 
\citep{nokafo2014}.

There is an alternative explanation for the faint X-ray outburst of
the source: the cooling of the crust of a Galactic neutron star heated
in a large accretion outburst. The main problem with this
explanation is that the accretion outburst was not detected by the
All-sky Monitor onboard \emph{RXTE} and would therefore be too weak to heat
up the crust of the neutron star (Lin18).

In order to differentiate the above two explanations, we obtained
follow-up observations with \emph{XMM-Newton} and \emph{HST} in
2018. The \emph{XMM-Newton} observation served to monitor the X-ray flux
and spectral evolution and to check whether the luminosity continues
to decrease as expected for a TDE. The \emph{HST} image served to check
whether the optical counterpart is extended or not.  In this Letter we
report the results of these new observations. In
Section~\ref{sec:reduction}, we describe the data analysis. In
Section~\ref{sec:res}, we present the results. The conclusions and the
discussion of the source nature are given in
Section~\ref{sec:conclusion}.

\begin{figure} 
\centering
\includegraphics[width=3.4in]{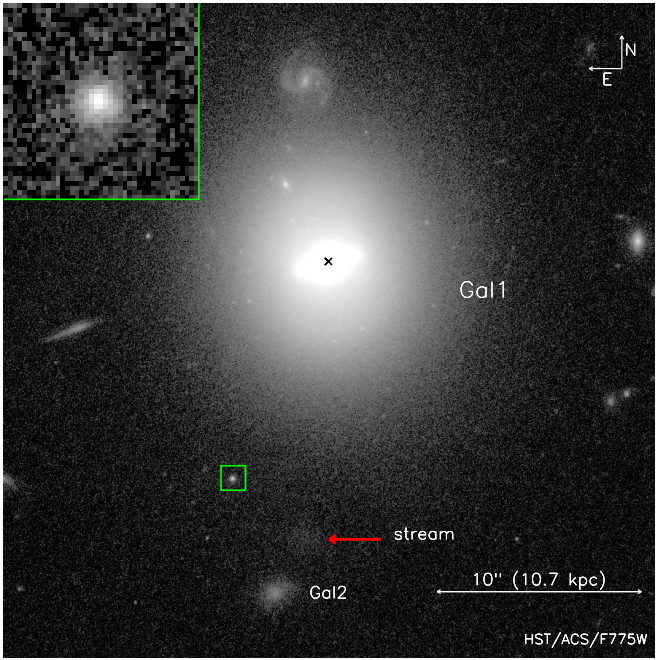}
\caption{The new \emph{HST} image around the field of
  J2150$-$0551. The green box of 1\farcs2$\times$1\farcs2, with zoomed
  inset, is centered around the
  source and is the region that we used to carry out the profile fitting
  (Section~\ref{sec:res}). Gal1 is the main host galaxy of the
  source, and near the source is a possible satellite galaxy Gal2,
  which might be connected with Gal1 by a tidal stream. \label{fig:hstimg2018}}
\end{figure}

\begin{figure} 
\centering
\includegraphics[width=3.4in]{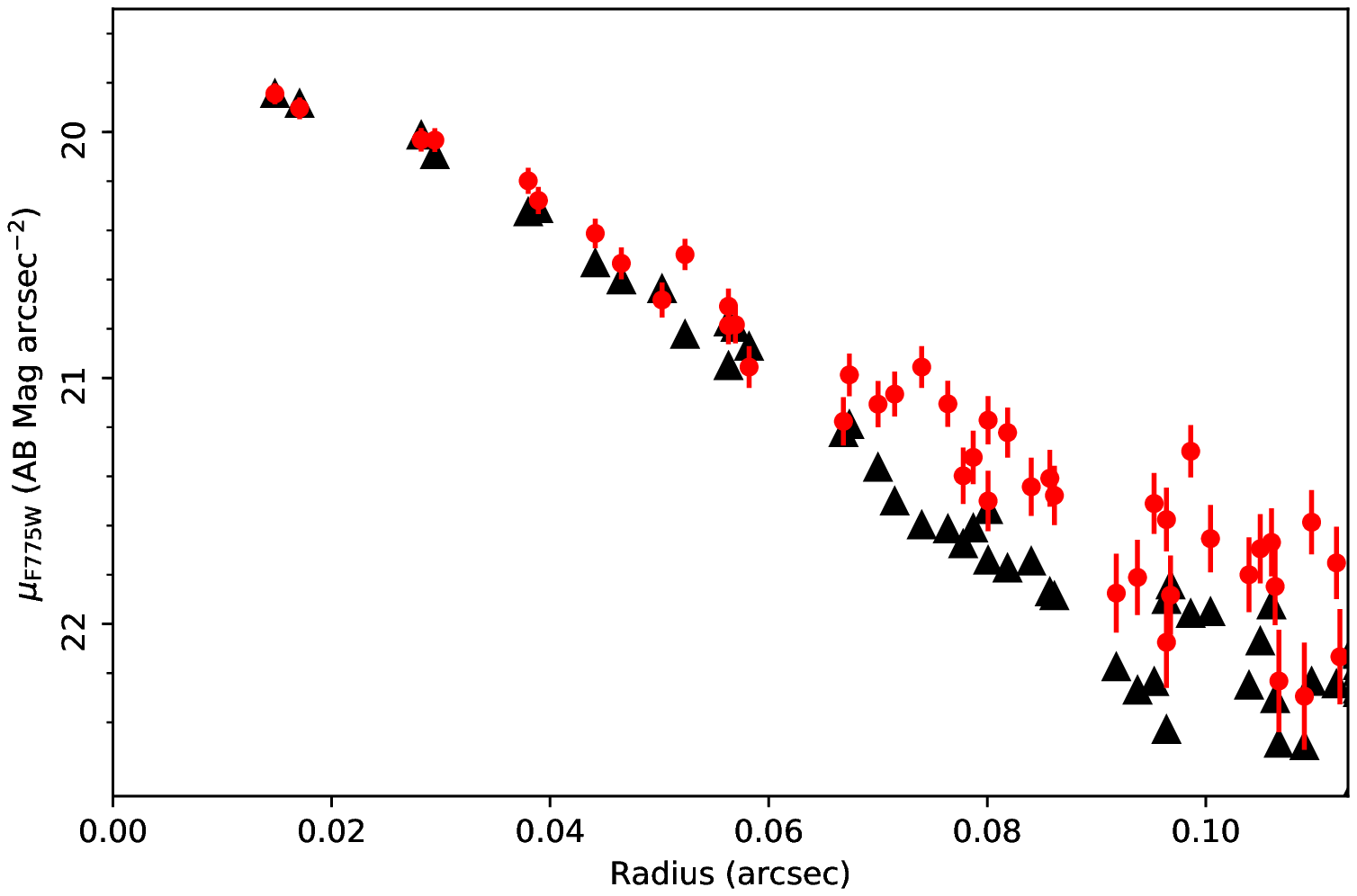}
\includegraphics[width=3.4in]{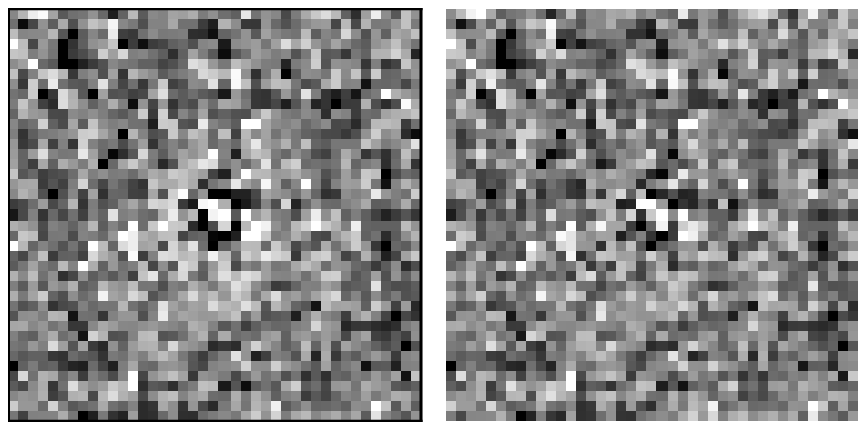}
\caption{Top panel: the surface brightness of the star cluster in different pixels versus
  their distances to the derived center (red circle, $1\sigma$ error). The black
  triangles (errors are not shown but are negligible) are for the empirical PSF image (derived from nearby stars) with the center and the
  peak value forced to
  align with those of  J2150$-$0551. Bottom panels: The
  two-dimensional residuals of the single S\'{e}rsic fits with index
  1.0 (left) and 6.0 (right), using the GALFIT software.\label{fig:psffit}}
\end{figure}

\begin{figure*} 
\centering
\includegraphics[width=5.0in]{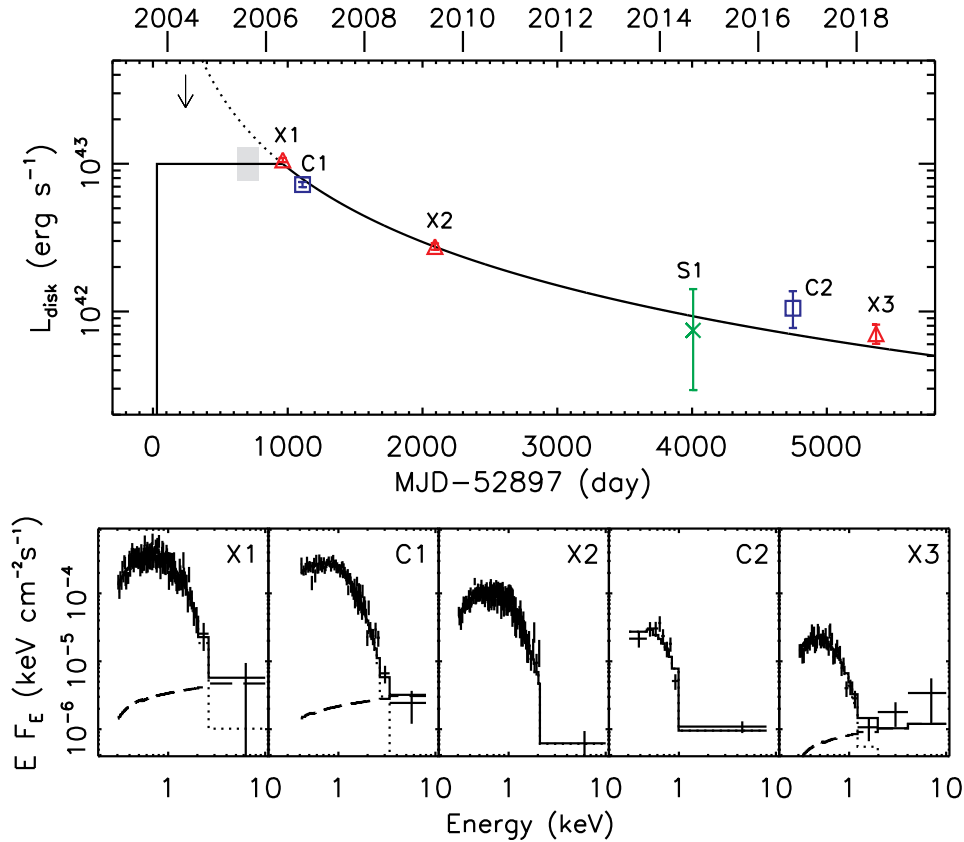}
\caption{Top panel: the long-term bolometric disk luminosity curve
  with 90\% errors
  from various pointed observations (\emph{Chandra} blue squares,
  \emph{XMM-Newton} red triangles, and \emph{Swift} green cross). The
  downward arrow in 2004 marks the $3\sigma$ upper limit from an
  \emph{XMM-Newton} slew observation and the gray shaded region marks
  the time interval when the optical flare was detected in 2005. The
 solid line is a simple TDE model, in which a very fast rise occurred
 one month after disruption and then the luminosity remained constant due to super-Eddington accretion effects (such as photon trapping),
 followed by a standard $t^{-5/3}$ decay. The dotted line neglects the
 super-Eddington accretion effects. Bottom panels: the standard thermal disk fits to the
 high-quality X-ray spectra at different epochs. A weak power-law was
 added  to account
 for possible contamination from the nuclear emission of Gal1 in
 all fits except C2 (the nuclear source is spatially separated in this observation). For visual purposes, the spectra are rebinned to
 be above $2\sigma$ in each bin in the plot. For \emph{XMM-Newton}
 observations, only pn spectra are shown. The plot is the same as
 Figure 2 in Lin18 except that we have added the new \emph{XMM-Newton}
 observation X3 and have updated the TDE model in the upper panel. \label{fig:ltlumsp}}
\end{figure*}

\section{DATA ANALYSIS}
\label{sec:reduction}
There are various multiwavelength observations of J2150$-$0551. 
In this Letter, we will focus on the new \emph{XMM-Newton} and
\emph{HST} observations obtained in 2018, though we will also include some results
of previous observations as obtained in Lin18. The new
\emph{XMM-Newton} observation (ObsID: 0823360101, X3 hereafter) was taken on 2018 May 24
in the imaging mode, with the exposure times of
49.3 ks, 57.8 ks, and 57.9 ks for the three European
Photon Imaging Cameras (EPIC) pn, MOS1, and
  MOS2, respectively. We used SAS 16.0.0 and the
calibration files of 2017 March as adopted in Lin18 for reprocessing the X-ray event
files and follow-up analysis. There are no clear background flares
seen in all cameras, and we used all data. We reprocessed the data and
extracted the source light curve and spectra in the standard way, and
we refer to Lin18 for details. Because the source was faint in X3, we
adopt a circular source region of radius 20 arcsec.

The new \emph{HST} observation was also carried out on 2018 May 24
under the program GO-15441,
and it was to obtain an ACS/WFC image with the F775W filter, as
adopted in the previous observation in 2003. It was composed of four
exposures of 544 s each (2176 s in total). We produced the drizzled
count image with the DrizzlePac software, with the pixel size set to
be 0.03 arcsec. We performed a profile fit to the counterpart to
J2150$-$0551 using two packages: one is GALFIT \citep{pehoim2010} and
the other is ISHAPE \citep{la1999}. An empirical point-spread function
(PSF) was derived from four nearby stars, with an oversampling factor
of ten.

\begin{figure}
\centering
\includegraphics[width=3.4in]{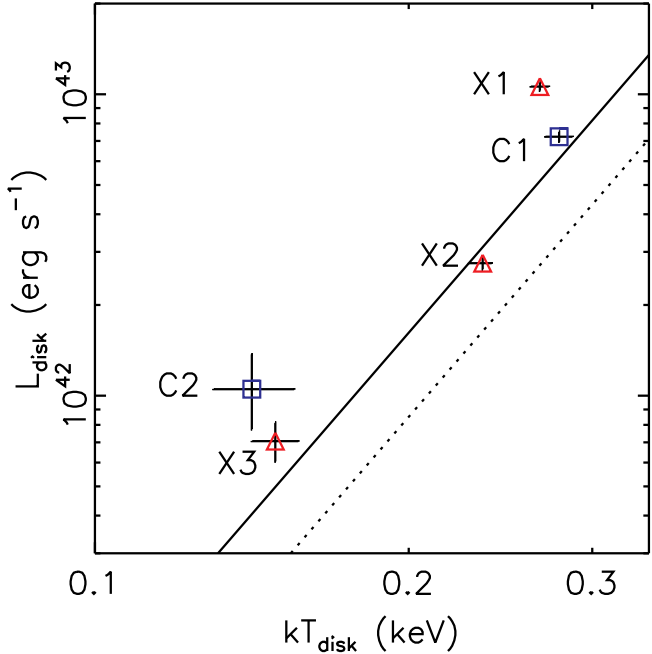}
\caption{The disk luminosity versus the apparent maximum
  temperature, with 90\% errors. The solid line plots the $L\propto T^4$ relation with
  the inner disk radius being the mean value of C1, X2, C2, and X3,
  weighted by the errors, while the dotted line plots the same
  relation but for ESO 243$-$49 HLX-1 \citep{sefali2011}. \label{fig:fluxtbb}}
\end{figure}

\section{RESULTS}
\label{sec:res}
 
\subsection{Deep Optical Imaging}
Figure~\ref{fig:hstimg2018} shows the new \emph{HST} ACS F775W image
around J2150$-$0551. The quality of the image around the source is
significantly improved compared with the one obtained in 2003, due to
twice the exposure length and better cosmic ray rejection (note the
location of the source in the CCD gap in two previous exposures).  We
measured the magnitude of $\mathrm{F775W}=24.08\pm0.01$ mag (AB, 1$\sigma$ error) and $24.02\pm0.02$
mag from the 2018 and 2003 images, respectively, using an aperture
of radius 0.3 arcsec (the background was estimated from four nearby
circular regions of the same size). Therefore, we observed no clear
optical variability between these two epochs. Because the 2003 image
was taken before the outburst and thus represents the
quiescence emission level (Lin18), the optical outburst, which was detected in
2005, must have subsided to below the detection level in 2018.

The counterpart seems marginally resolved in the 2018 image. This can
be seen from the comparison of its radial profile with that of the PSF
shown in the top panel in Figure~\ref{fig:psffit}. 

We first used GALFIT to fit the optical profile of the source in the
new \emph{HST} image with a single S\'{e}rsic function (convolved with the
PSF). At the position of the source, there is star light from the host
galaxy Gal1. Then there are also concerns about the size of the fitting region
and how to model the background (sky plus the star light from
Gal1). We first tried a fitting region of 1\farcs2$\times$1\farcs2
centered around the source. The background was allowed to be variable
and have a linear gradient. The profile seems symmetric, and the fit inferred the
axis ratio to be consistent with 1.0. Therefore we fixed the value of
this parameter at 1.0.  We found that the fits preferred a large
S\'{e}rsic index. The bottom panels of Figure~\ref{fig:psffit} compare
the fit residuals of index fixed at 1.0 (left) and those of index fixed at
6.0 (right). The fit with index 6.0 is reasonably acceptable (reduced
$\chi^2$ value of $\chi^2_\nu=0.920$ for $\nu=1675$ degrees of freedom)  and is better than that of index 1.0, with the
total $\chi^2$ reduced by 94.0. Assuming a higher value of index would
reduce the $\chi^2$ value further but very slightly. The inferred
half-light radius is not sensitive to the index value assumed and was
inferred to be around 0.024--0.026 arcsec (i.e. 26--28 pc). The
integrated magnitude is fainter for smaller index ($\mathrm{F775W}=24.13$ mag for
index 1.0 and 23.92 mag for index 6.0). We note that the fits do not
require a background gradient, which means that the fitting region
used is not too large. The inferred background value from all
fits is consistent with the median from an annulus of inner and outer
radii of 0.5 arcsec and 0.7 arcsec, respectively.

There is no significant improvement to the fit using two S\'{e}rsic
functions, reducing the $\chi^2$ value by only 20, compared with the
single S\'{e}rsic fit with index 6.0. The parameters could not be
constrained well, and one possible good fit could be two S\'{e}rsic
functions, with index 1.0 and 4.0 (fixed), effective radius 140 pc and
5 pc, and integrated magnitude $\mathrm{F775W}=25.01$ mag and 24.37 mag,
respectively. 

We also tested other popular models that are often used to fit star
clusters. With a Moffat model of index 1.5, we obtained a reasonable
fit ($\chi^2_\nu=0.935$) with an effective radius of 27 pc. Using a King model, we obtained
reasonable fits ($\chi^2_\nu\sim0.92$) with large concentration values and effective radius
$\sim$23 pc. These fits suggest that although there are multiple
models that can fit the optical counterpart to J2150$-$0551 well, the
inferred effective radius is fairly consistent among different models.

We also used ISHAPE to carry out fits. We obtained
consistent results with those with GALFIT. ISHAPE inferred the effective radius from
various models to range between 26--32 pc.

\subsection{X-ray follow-up}
The X-ray spectrum of J2150$-$0551 from X3 is very soft, with most
emission from low-energy photons below 1 keV
(Figure~\ref{fig:ltlumsp}). Therefore, the source should be still in
the thermal state, and we fit the spectrum with an absorbed standard
thermal disk model (\emph{diskbb} in XSPEC)  as we did for previous observations (Lin18). A very
weak powerlaw of a photon index fixed at 1.8 was included to account
for contamination of the nuclear emission from Gal1, as was found in
the high-resolution \emph{Chandra} observation in 2016 (C2, Lin18). We
inferred the disk apparent maximum temperature to be $0.149\pm0.008$
keV and the disk bolometric luminosity to be
$7\pm1$$\times10^{41}$ erg
s$^{-1}$ (errors are 90\%). The luminosity is 30\%
lower than that from the previous \emph{Chandra} observation C2, but
such a decrease is not significant due to the large error bar of the
\emph{Chandra} observation (Figure~\ref{fig:ltlumsp}). Lin18
constructed a simple TDE model to explain the luminosity evolution of
the event. The updated model from the inclusion of the new additional
observation X3 is shown in Figure~\ref{fig:ltlumsp} (solid line in the
upper panel). The disruption time was inferred to be on 2003 September
15 (close to the time 2003 October 18 found in Lin18), with
1$\sigma$ uncertainty of 35 days. The luminosity in X3 is close to
that predicted in the simple TDE model (within 1.7$\sigma$). According
to the model, the total energy radiated until X3 was $1.7\times10^{51}$
ergs, and the total mass accreted into the black hole until X3 was
$0.069 (0.1/\eta)$ \msun, where $\eta$ is the rest mass to raditation energy
conversion efficiency in the sub-Eddington accretion phase.

The updated disk luminosity versus disk apparent maximum temperature
plot including X3 is shown in Figure~\ref{fig:fluxtbb}. The new
observation is consistent (to within 1.7$\sigma$) with the $L\propto
T^4$ relation traced out by C1, X2, and C2. This strongly supports
that the source was still in the thermal state in X3.

\section{DISCUSSION AND CONCLUSIONS}
\label{sec:conclusion}
The most significant result of our multiwavelength follow-up
observations is the confirmation of the optical counterpart to
J2150$-$0551 as an extended but very compact source, when the optical emission associated with the X-ray activity has
subsided to below detection level. The
extended nature of the counterpart firmly rules out the Galactic
cooling neutron star explanation for the source, which is the reason why
we did not test the neutron star atmosphere model on X3 as we did for
previous X-ray spectra in Lin18. Therefore,  J2150$-$0551 should have
an extragalactic origin and can be associated with either Gal1 or a
background galaxy.  The chance probability for J2150$-$0551 to be
within 11.6 arcsec from the center of a bright galaxy like Gal1 is
very small, only 0.01\% (Lin18). The optical counterpart to
J2150$-$0551 has a circularly symmetric compact profile, unlike most
background galaxies. The high disk temperature of the outburst is also hard to explain if it is in a
distant background galaxy, as will be discussed below. Therefore,
J2150$-$0551 is most likely associated with Gal1.

With a half-light radius of $\sim$27 pc, an absolute $V$-band magnitude
of $-12.3$ AB mag and a stellar mass of $\sim$$10^7$ \msun, the
counterpart could be a massive globular cluster or a UCD resulting from
a minor merger \citep{nokafo2014}. The latter explanation is more likely, given that
the host galaxy Gal1 seems to be in an epoch of active minor mergers
(see the presence of a possible nearby minor merger of Gal2 with Gal1,
Figure~\ref{fig:hstimg2018}).  This compact system has a stellar mass around
the limit above which stellar systems could only be explained as
remnant nuclei of tidally disrupted galaxies, instead of true ancient
star clusters \citep{novaka2019}.  

The other main result that we obtained is the continuing decay of the
source luminosity according to a simple TDE evolution model. The
 X-ray light curve spans 12 years. In addition, the
X-ray spectrum remained supersoft, with good statistics confirming the
disk luminosity evolution following
the $L\propto T^4$ relation. This is a smoking-gun evidence for the
thermal state of an accreting BH. Although the $L\propto T^4$ relation
is commonly seen in BH X-ray binaries, such a relation was only
obtained in a few cases for accreting massive BHs, especially TDEs
associated with nuclear SMBHs. Because the disk
temperature depends on the BH mass as $M_{\rm BH}^{-1/4}$ for a given
Eddington ratio, it is expected that IMBHs have higher disk temperatures
than SMBHs. This explains why the IMBH candidates J2150$-$0551 and ESO
243$-$49 HLX-1 \citep{sefali2011,goplka2012} have disk temperatures reaching $\sim$0.25 keV, much
higher than observed in SMBH TDEs that also exhibited $L\propto T^4$ \citep[$\lesssim$0.1 keV;][]{licagr2011,misagi2019}.

\acknowledgments Acknowledgments: This work is supported by  the
National Aeronautics and Space Administration 
XMM-Newton GO program grant
80NSSC19K0873, by National Aeronautics and Space Administration
through grant number HST-GO-15441.001-A from the Space Telescope Science Institute, which is operated
by AURA, Inc., under NASA contract NAS 5-26555, and by the National Aeronautics and Space
Administration ADAP grant NNX17AJ57G. JS acknowledges support from the
Packard Foundation. AJR was supported as a Research
Corporation for Science Advancement Cottrell Scholar. NW, OG and DB acknowledge CNES for financial support to the XMM-Newton Survey Science Center activities.


\begin{thebibliography}{32}
\expandafter\ifx\csname natexlab\endcsname\relax\def\natexlab#1{#1}\fi

\bibitem[{{Abbott} {et~al.}(2016){Abbott}, {Abbott}, {Abbott}, {Abernathy},
  {Acernese}, {Ackley}, {Adams}, {Adams}, {Addesso}, {Adhikari}, \&
  et~al.}]{ababab2016}
{Abbott}, B.~P., {Abbott}, R., {Abbott}, T.~D., {et~al.} 2016, Physical Review
  Letters, 116, 061102

\bibitem[{{Baldassare} {et~al.}(2015){Baldassare}, {Reines}, {Gallo}, \&
  {Greene}}]{barega2015}
{Baldassare}, V.~F., {Reines}, A.~E., {Gallo}, E., \& {Greene}, J.~E. 2015,
  \apjl, 809, L14
  
\bibitem[{{Baumgardt} {et~al.}(2019){Baumgardt}, {He}, {Sweet}, {Drinkwater},
  {Sollima}, {Hurley}, {Usher}, {Kamann}, {Dalgleish}, {Dreizler}, \&
  {Husser}}]{bahesw2019}
{Baumgardt}, H., {He}, C., {Sweet}, S.~M., {et~al.} 2019, \mnras, 488, 5340

\bibitem[{{Chen} \& {Shen}(2018)}]{chsh2018}
{Chen}, J.-H. \& {Shen}, R.-F. 2018, \apj, 867, 20

\bibitem[{{Chilingarian} {et~al.}(2018){Chilingarian}, {Katkov}, {Zolotukhin},
  {Grishin}, {Beletsky}, {Boutsia}, \& {Osip}}]{chkazo2018}
{Chilingarian}, I.~V., {Katkov}, I.~Y., {Zolotukhin}, I.~Y., {et~al.} 2018,
  \apj, 863, 1

\bibitem[{{Dong} {et~al.}(2007){Dong}, {Wang}, {Yuan}, {Shan}, {Zhou}, {Fan},
  {Dou}, {Wang}, {Wang}, \& {Lu}}]{dowayu2007}
{Dong}, X., {Wang}, T., {Yuan}, W., {et~al.} 2007, \apj, 657, 700

\bibitem[{{Event Horizon Telescope Collaboration} {et~al.}(2019){Event Horizon
  Telescope Collaboration}, {Akiyama}, {Alberdi}, {Alef}, {Asada}, {Azulay},
  {Baczko}, {Ball}, {Balokovi{\'c}}, {Barrett}, {Bintley}, {Blackburn},
  {Boland}, {Bouman}, {Bower}, {Bremer}, {Brinkerink}, {Brissenden}, {Britzen},
  {Broderick}, {Broguiere}, {Bronzwaer}, {Byun}, {Carlstrom}, {Chael}, {Chan},
  {Chatterjee}, {Chatterjee}, {Chen}, {Chen}, {Cho}, {Christian}, {Conway},
  {Cordes}, {Crew}, {Cui}, {Davelaar}, {De Laurentis}, {Deane}, {Dempsey},
  {Desvignes}, {Dexter}, {Doeleman}, {Eatough}, {Falcke}, {Fish}, {Fomalont},
  {Fraga-Encinas}, {Freeman}, {Friberg}, {Fromm}, {G{\'o}mez}, {Galison},
  {Gammie}, {Garc{\'\i}a}, {Gentaz}, {Georgiev}, {Goddi}, {Gold}, {Gu},
  {Gurwell}, {Hada}, {Hecht}, {Hesper}, {Ho}, {Ho}, {Honma}, {Huang}, {Huang},
  {Hughes}, {Ikeda}, {Inoue}, {Issaoun}, {James}, {Jannuzi}, {Janssen},
  {Jeter}, {Jiang}, {Johnson}, {Jorstad}, {Jung}, {Karami}, {Karuppusamy},
  {Kawashima}, {Keating}, {Kettenis}, {Kim}, {Kim}, {Kim}, {Kino}, {Koay},
  {Koch}, {Koyama}, {Kramer}, {Kramer}, {Krichbaum}, {Kuo}, {Lauer}, {Lee},
  {Li}, {Li}, {Lindqvist}, {Liu}, {Liuzzo}, {Lo}, {Lobanov}, {Loinard},
  {Lonsdale}, {Lu}, {MacDonald}, {Mao}, {Markoff}, {Marrone}, {Marscher},
  {Mart{\'\i}-Vidal}, {Matsushita}, {Matthews}, {Medeiros}, {Menten}, {Mizuno},
  {Mizuno}, {Moran}, {Moriyama}, {Moscibrodzka}, {M{\"u}ller}, {Nagai},
  {Nagar}, {Nakamura}, {Narayan}, {Narayanan}, {Natarajan}, {Neri}, {Ni},
  {Noutsos}, {Okino}, {Olivares}, {Ortiz-Le{\'o}n}, {Oyama}, {{\"O}zel},
  {Palumbo}, {Patel}, {Pen}, {Pesce}, {Pi{\'e}tu}, {Plambeck}, {PopStefanija},
  {Porth}, {Prather}, {Preciado-L{\'o}pez}, {Psaltis}, {Pu}, {Ramakrishnan},
  {Rao}, {Rawlings}, {Raymond}, {Rezzolla}, {Ripperda}, {Roelofs}, {Rogers},
  {Ros}, {Rose}, {Roshanineshat}, {Rottmann}, {Roy}, {Ruszczyk}, {Ryan},
  {Rygl}, {S{\'a}nchez}, {S{\'a}nchez-Arguelles}, {Sasada}, {Savolainen},
  {Schloerb}, {Schuster}, {Shao}, {Shen}, {Small}, {Sohn}, {SooHoo}, {Tazaki},
  {Tiede}, {Tilanus}, {Titus}, {Toma}, {Torne}, {Trent}, {Trippe}, {Tsuda},
  {van Bemmel}, {van Langevelde}, {van Rossum}, {Wagner}, {Wardle},
  {Weintroub}, {Wex}, {Wharton}, {Wielgus}, {Wong}, {Wu}, {Young}, {Young},
  {Younsi}, {Yuan}, {Yuan}, {Zensus}, {Zhao}, {Zhao}, {Zhu}, {Algaba},
  {Allardi}, {Amestica}, {Anczarski}, {Bach}, {Baganoff}, {Beaudoin}, {Benson},
  {Berthold}, {Blanchard}, {Blundell}, {Bustamente}, {Cappallo},
  {Castillo-Dom{\'\i}nguez}, {Chang}, {Chang}, {Chang}, {Chen}, {Chilson},
  {Chuter}, {C{\'o}rdova Rosado}, {Coulson}, {Crawford}, {Crowley}, {David},
  {Derome}, {Dexter}, {Dornbusch}, {Dudevoir}, {Dzib}, {Eckart}, {Eckert},
  {Erickson}, {Everett}, {Faber}, {Farah}, {Fath}, {Folkers}, {Forbes},
  {Freund}, {G{\'o}mez-Ruiz}, {Gale}, {Gao}, {Geertsema}, {Graham}, {Greer},
  {Grosslein}, {Gueth}, {Haggard}, {Halverson}, {Han}, {Han}, {Hao},
  {Hasegawa}, {Henning}, {Hern{\'a}ndez-G{\'o}mez}, {Herrero-Illana},
  {Heyminck}, {Hirota}, {Hoge}, {Huang}, {Impellizzeri}, {Jiang}, {Kamble},
  {Keisler}, {Kimura}, {Kono}, {Kubo}, {Kuroda}, {Lacasse}, {Laing}, {Leitch},
  {Li}, {Lin}, {Liu}, {Liu}, {Lu}, {Marson}, {Martin-Cocher}, {Massingill},
  {Matulonis}, {McColl}, {McWhirter}, {Messias}, {Meyer-Zhao}, {Michalik},
  {Monta{\~n}a}, {Montgomerie}, {Mora-Klein}, {Muders}, {Nadolski}, {Navarro},
  {Neilsen}, {Nguyen}, {Nishioka}, {Norton}, {Nowak}, {Nystrom}, {Ogawa},
  {Oshiro}, {Oyama}, {Parsons}, {Paine}, {Pe{\~n}alver}, {Phillips}, {Poirier},
  {Pradel}, {Primiani}, {Raffin}, {Rahlin}, {Reiland}, {Risacher}, {Ruiz},
  {S{\'a}ez-Mada{\'\i}n}, {Sassella}, {Schellart}, {Shaw}, {Silva}, {Shiokawa},
  {Smith}, {Snow}, {Souccar}, {Sousa}, {Sridharan}, {Srinivasan}, {Stahm},
  {Stark}, {Story}, {Timmer}, {Vertatschitsch}, {Walther}, {Wei}, {Whitehorn},
  {Whitney}, {Woody}, {Wouterloot}, {Wright}, {Yamaguchi}, {Yu}, {Zeballos},
  {Zhang}, \& {Ziurys}}]{evakal2019}
{Event Horizon Telescope Collaboration}, {Akiyama}, K., {Alberdi}, A., {et~al.}
  2019, \apjl, 875, L1

\bibitem[{{Farrell} {et~al.}(2009){Farrell}, {Webb}, {Barret}, {Godet}, \&
  {Rodrigues}}]{faweba2009}
{Farrell}, S.~A., {Webb}, N.~A., {Barret}, D., {Godet}, O., \& {Rodrigues},
J.~M. 2009, \nat, 460, 73

\bibitem[{{Gillessen} {et~al.}(2009){Gillessen}, {Eisenhauer}, {Trippe},
  {Alexand er}, {Genzel}, {Martins}, \& {Ott}}]{gieitr2009}
{Gillessen}, S., {Eisenhauer}, F., {Trippe}, S., {et~al.} 2009, \apj, 692, 1075

\bibitem[{{Godet} {et~al.}(2012){Godet}, {Plazolles}, {Kawaguchi}, {Lasota},
  {Barret}, {Farrell}, {Braito}, {Servillat}, {Webb}, \&
  {Gehrels}}]{goplka2012}
{Godet}, O., {Plazolles}, B., {Kawaguchi}, T., {et~al.} 2012, \apj, 752, 34

\bibitem[{{Gravity Collaboration} {et~al.}(2018){Gravity Collaboration},
  {Abuter}, {Amorim}, {Baub{\"o}ck}, {Berger}, {Bonnet}, {Brand ner},
  {Cl{\'e}net}, {Coud{\'e} Du Foresto}, {de Zeeuw}, {Deen}, {Dexter}, {Duvert},
  {Eckart}, {Eisenhauer}, {F{\"o}rster Schreiber}, {Garcia}, {Gao}, {Gendron},
  {Genzel}, {Gillessen}, {Guajardo}, {Habibi}, {Haubois}, {Henning}, {Hippler},
  {Horrobin}, {Huber}, {Jim{\'e}nez-Rosales}, {Jocou}, {Kervella}, {Lacour},
  {Lapeyr{\`e}re}, {Lazareff}, {Le Bouquin}, {L{\'e}na}, {Lippa}, {Ott},
  {Panduro}, {Paumard}, {Perraut}, {Perrin}, {Pfuhl}, {Plewa}, {Rabien},
  {Rodr{\'\i}guez-Coira}, {Rousset}, {Sternberg}, {Straub}, {Straubmeier},
  {Sturm}, {Tacconi}, {Vincent}, {von Fellenberg}, {Waisberg}, {Widmann},
  {Wieprecht}, {Wiezorrek}, {Woillez}, \& {Yazici}}]{grabam2018}
{Gravity Collaboration}, {Abuter}, R., {Amorim}, A., {et~al.} 2018, \aap, 618,
  L10

\bibitem[{{Greene} {et~al.}(2019){Greene}, {Strader}, \& {Ho}}]{grstho2019}
{Greene}, J.~E., {Strader}, J., \& {Ho}, L.~C. 2019, arXiv:1911.09678

\bibitem[{{Irwin} {et~al.}(2010){Irwin}, {Brink}, {Bregman}, \&
  {Roberts}}]{irbrbr2010}
{Irwin}, J.~A., {Brink}, T.~G., {Bregman}, J.~N., \& {Roberts}, T.~P. 2010,
  \apjl, 712, L1

\bibitem[{{Israel} {et~al.}(2017){Israel}, {Belfiore}, {Stella}, {Esposito},
  {Casella}, {De Luca}, {Marelli}, {Papitto}, {Perri}, {Puccetti}, {Castillo},
  {Salvetti}, {Tiengo}, {Zampieri}, {Agostino}, {Greiner}, {Haberl}, {Novara},
  {Salvaterra}, {Turolla}, {Watson}, {Wilms}, \& {Wolter}}]{isbest2017}
{Israel}, G.~L., {Belfiore}, A., {Stella}, L., {et~al.} 2017, Science, 355, 817

\bibitem[{{K{\i}z{\i}ltan} {et~al.}(2017){K{\i}z{\i}ltan}, {Baumgardt}, \&
  {Loeb}}]{kibalo2017}
{K{\i}z{\i}ltan}, B., {Baumgardt}, H., \& {Loeb}, A. 2017, \nat, 542, 203

\bibitem[{{Komossa}(2015)}]{ko2015}
{Komossa}, S. 2015, JHEA, 7, 148

\bibitem[{{Kormendy} \& {Richstone}(1995)}]{kori1995}
{Kormendy}, J. \& {Richstone}, D. 1995, \araa, 33, 581

\bibitem[{{Larsen}(1999)}]{la1999}
{Larsen}, S.~S. 1999, \aaps, 139, 393

\bibitem[{{Lin} {et~al.}(2011){Lin}, {Carrasco}, {Grupe}, {Webb}, {Barret}, \&
  {Farrell}}]{licagr2011}
{Lin}, D., {Carrasco}, E.~R., {Grupe}, D., {et~al.} 2011, \apj, 738, 52

\bibitem[{{Lin} {et~al.}(2016){Lin}, {Carrasco}, {Webb}, {Irwin}, {Dupke},
  {Romanowsky}, {Ramirez-Ruiz}, {Strader}, {Homan}, {Barret}, \&
  {Godet}}]{licawe2016}
{Lin}, D., {Carrasco}, E.~R., {Webb}, N.~A., {et~al.} 2016, \apj, 821, 25

\bibitem[{{Lin} {et~al.}(2018){Lin}, {Strader}, {Carrasco}, {Page},
  {Romanowsky}, {Homan}, {Irwin}, {Remillard}, {Godet}, {Webb}, {Baumgardt},
  {Wijnands}, {Barret}, {Duc}, {Brodie}, \& {Gwyn}}]{listca2018}
{Lin}, D., {Strader}, J., {Carrasco}, E.~R., {et~al.} 2018, Nature Astronomy,
  2, 656

\bibitem[{{Mann} {et~al.}(2019){Mann}, {Richer}, {Heyl}, {Anderson}, {Kalirai},
  {Caiazzo}, {M{\"o}hle}, {Knee}, \& {Baumgardt}}]{marihe2019}
{Mann}, C.~R., {Richer}, H., {Heyl}, J., {et~al.} 2019, \apj, 875, 1

\bibitem[{{Miniutti} {et~al.}(2019){Miniutti}, {Saxton}, {Giustini},
  {Alexander}, {Fender}, {Heywood}, {Monageng}, {Coriat}, {Tzioumis}, {Read},
  {Knigge}, {Gandhi}, {Pretorius}, \& {Ag{\'{\i}}s-Gonz{\'a}lez}}]{misagi2019}
{Miniutti}, G., {Saxton}, R.~D., {Giustini}, M., {et~al.} 2019, \nat, 573, 381

\bibitem[{{Norris} {et~al.}(2014){Norris}, {Kannappan}, {Forbes}, {Romanowsky},
  {Brodie}, {Faifer}, {Huxor}, {Maraston}, {Moffett}, {Penny}, {Pota},
  {Smith-Castelli}, {Strader}, {Bradley}, {Eckert}, {Fohring}, {McBride},
  {Stark}, \& {Vaduvescu}}]{nokafo2014}
{Norris}, M.~A., {Kannappan}, S.~J., {Forbes}, D.~A., {et~al.} 2014, \mnras,
  443, 1151

\bibitem[{{Norris} {et~al.}(2019){Norris}, {van de Ven}, {Kannappan},
  {Schinnerer}, \& {Leaman}}]{novaka2019}
{Norris}, M.~A., {van de Ven}, G., {Kannappan}, S.~J., {Schinnerer}, E., \&
  {Leaman}, R. 2019, \mnras, 488, 5400

\bibitem[{{Noyola} {et~al.}(2008){Noyola}, {Gebhardt}, \&
  {Bergmann}}]{nogebe2008}
{Noyola}, E., {Gebhardt}, K., \& {Bergmann}, M. 2008, \apj, 676, 1008

\bibitem[{{Peng} {et~al.}(2010){Peng}, {Ho}, {Impey}, \& {Rix}}]{pehoim2010}
{Peng}, C.~Y., {Ho}, L.~C., {Impey}, C.~D., \& {Rix}, H.-W. 2010, \aj, 139,
  2097

\bibitem[{{Perera} {et~al.}(2017){Perera}, {Stappers}, {Lyne}, {Bassa},
  {Cognard}, {Guillemot}, {Kramer}, {Theureau}, \& {Desvignes}}]{pestly2017}
{Perera}, B.~B.~P., {Stappers}, B.~W., {Lyne}, A.~G., {et~al.} 2017, \mnras,
  468, 2114

\bibitem[{{Remillard} \& {McClintock}(2006)}]{remc2006}
{Remillard}, R.~A. \& {McClintock}, J.~E. 2006, \araa, 44, 49

\bibitem[{{Servillat} {et~al.}(2011){Servillat}, {Farrell}, {Lin}, {Godet},
  {Barret}, \& {Webb}}]{sefali2011}
{Servillat}, M., {Farrell}, S.~A., {Lin}, D., {et~al.} 2011, \apj, 743, 6

\bibitem[{{Sutton} {et~al.}(2015){Sutton}, {Roberts}, {Gladstone}, \&
  {Walton}}]{surogl2015}
{Sutton}, A.~D., {Roberts}, T.~P., {Gladstone}, J.~C., \& {Walton}, D.~J. 2015,
  \mnras, 450, 787

\bibitem[{{Webb} {et~al.}(2012){Webb}, {Cseh}, {Lenc}, {Godet}, {Barret},
  {Corbel}, {Farrell}, {Fender}, {Gehrels}, \& {Heywood}}]{wecsle2012}
{Webb}, N., {Cseh}, D., {Lenc}, E., {et~al.} 2012, Science, 337, 554

\end{thebibliography}
\end{document}